\documentclass[argument,twocolumn]{aastex62}
\usepackage{graphicx}	
\usepackage{subfigure}	
\usepackage{amsmath}	
\usepackage{amssymb}	
\usepackage{bm}
\usepackage{ulem}
\usepackage{color}

\newcommand{\hahb}{{{\rm H}\alpha/{\rm H}\beta}}
\newcommand{\rebv}{R_{\rm EBV}}
\newcommand{\ebvhahb}{{\rm E(B-V)}_{\rm \hahb}}
\newcommand{\ebvirx}{{\rm E(B-V)}_{\rm IRX}}
\newcommand{\hii}{H\,{\sc ii}}
\newcommand{\mussfr}{{\rm SSFR}/R_{\rm e}^{2}} 
\received{XXX}
\revised{XXX}
\accepted{XXX}
\submitjournal{ApJ}
\shorttitle{The discrepancy between IRX and Balmer decrement in SFGs }
\shortauthors{Qin et al.}

\begin{document}

\title{Understanding the Discrepancy between IRX and Balmer Decrement in Tracing Galaxy Dust Attenuation}

\correspondingauthor{Xian~Zhong Zheng}
\email{jbqin@pmo.ac.cn (JQ),  xzzheng@pmo.ac.cn (XZZ)}

\author[0000-0002-5179-1039]{Jianbo Qin}
\affiliation{Purple Mountain Observatory, Chinese Academy of Sciences, 10 Yuanhua Road, Nanjing 210033, China}
\affiliation{School of Astronomy and Space Sciences, University of Science and Technology of China, Hefei 230026, China}

\author[0000-0003-3728-9912]{Xian~Zhong Zheng}
\affiliation{Purple Mountain Observatory, Chinese Academy of Sciences, 10 Yuanhua Road, Nanjing 210033, China}

\author[0000-0003-3735-1931]{Stijn Wuyts}
\affiliation{Department of Physics, University of Bath, Claverton Down, Bath, BA2 7AY, UK}

\author[0000-0001-5662-8217]{Zhizheng Pan}
\affiliation{Purple Mountain Observatory, Chinese Academy of Sciences, 10 Yuanhua Road, Nanjing 210033, China}

\author{Jian Ren}
\affiliation{Purple Mountain Observatory, Chinese Academy of Sciences, 10 Yuanhua Road, Nanjing 210033, China}
\affiliation{School of Astronomy and Space Sciences, University of Science and Technology of China, Hefei 230026, China}

\begin{abstract}

We compare the infrared excess (IRX) and Balmer decrement ($\hahb$) as dust attenuation indicators in relation to other galaxy parameters using a sample of $\sim$32\,000 local star-forming galaxies (SFGs) carefully selected from SDSS, {\it GALEX} and {\it WISE}. While at fixed $\hahb$, IRX turns out to be independent on galaxy stellar mass, the Balmer decrement does show a strong mass dependence at fixed IRX.  We find the discrepancy, parameterized by the color excess ratio $\rebv \equiv E(B-V)_{\rm{IRX}}/E(B-V)_{\rm \hahb}$, is not dependent on the gas-phase metallicity and axial ratio but on the specific star formation rate (SSFR) and galaxy size ($R_{\rm e}$) following $\rebv=0.79+0.15\log(\mussfr)$. This finding reveals that the nebular attenuation as probed by the Balmer decrement becomes increasingly larger than the global (stellar) attenuation of SFGs with decreasing SSFR surface density. This can be understood in the context of an enhanced fraction of intermediate-age stellar populations that are less attenuated by dust than the \hii\ region-traced young population, in conjunction with a decreasing dust opacity of the diffuse ISM when spreading over a larger spatial extent.  Once the SSFR surface density of an SFG is known, the conversion between attenuation of nebular and stellar emission can be well estimated using our scaling relation.

\end{abstract}
\keywords{dust, extinction --- Galaxies: evolution --- Galaxies: ISM --- Galaxies: star formation }

\section{Introduction} \label{sec:sec1}

Dust accounts for only a small fraction ($\sim$0.1\%) of the baryonic mass in star-forming galaxies (SFGs), but plays a central role in transforming the ultraviolet (UV) and optical radiation into the far-infrared (IR) and attenuates the galaxy observables \citep[e.g.,][and references therein]{Galliano2018}. The global dust attenuation of a galaxy  relies on not only the dust content, the grain size distribution and on chemical composition, but also on the geometrical distribution of dust and stars in the galaxy \citep{Wild2011,Price2014,Reddy2015,Popping2017}. Characterizing dust attenuation is thus critical to deriving the physical properties of galaxies, and understanding the dust enrichment in line with galaxy evolution.

In practice, the UV slope, Balmer decrement ($\hahb$) and infrared excess (IRX: IR-to-UV luminosity ratio) are well-known probes for dust attenuation.\footnote{Here dust attenuation is also referred to as dust obscuration throughout this work.}  It has become clear that the UV slope is so sensitive to dust attenuation that it often probes the attenuation in the surface regions of heavily-obscured galaxies \citep[i.e., ``skin-effect'';][]{Popping2017,Wang2018}; and it also suffers from systematic uncertainties that couple with the stellar population age  \citep{Kong2004,Grasha2013,Popping2017,Qiu2019}. On the other hand, both Balmer decrement and IRX are barely affected by the ``skin effect'' and thus more often used as tracers of dust attenuation \citep[e.g.,][]{Wang1996,Martin2005,Garn2010,Xiao2012,Yuan2018,Koyama2019,Li2019,Qin2019}. 

However, estimates of dust attenuation based on Balmer decrement and IRX often yield inconsistent results. \citet{Garn2010} found the stellar mass is the most fundamental parameter in regulating $\hahb$ \citep[see also][]{Zahid2017}. In contrast, IRX is found to be determined by IR luminosity ($L_{\rm IR}$), metallicity ($Z$), galaxy size ($R_{\rm e}$) and axial ratio (b/a), with no dependence on stellar mass ($M_\ast$) \citep{Qin2019}. By examining the IRX--$\hahb$ (converted to E(B-V)$_{\rm star}$ and E(B-V)$_{\rm gas}$ respectively) diagram color-coded with stellar masses, \citet{Koyama2019} denoted that Balmer decrement increases with stellar mass at a fixed IRX, while IRX shows no dependence on stellar mass if Balmer decrement is fixed, suggestive of the ``extra attenuation'' (parameterized by the $\hahb$-to-IRX color excess ratio) that is correlated with stellar mass. They also found that the ``extra attenuation'' is correlated with star formation rate (SFR) or specific SFR (SSFR). Although IRX and Balmer decrement might probe intrinsically different levels of attenuation on the stars and nebulae in galaxies \citep[see][for discussions]{Koyama2019}, the discrepancy between the two remains to be understood.

In this work, we re-examine the discrepancy of dust attenuation between IRX and Balmer decrement in local SFGs to see what lead to the difference. In Section~\ref{sec:sec2}, we present a brief description of the galaxy sample and data. Section~\ref{sec:sec3} shows our results how the $\hahb$--IRX relation is affected by the different galaxy parameters. We discuss our results in Section~\ref{sec:sec4} and give a summary in Section~\ref{sec:sec5}. A standard $\Lambda$CDM cosmology with $H_0$=70\,km$^{-1}$\,Mpc$^{-1}$, $\Omega _{\rm \Lambda} = 0.7$ and $\Omega _{\rm m} = 0.3$ and a \citet{Chabrier2003} Initial Mass Function are adopted throughout the paper.

\begin{figure*}
\centering
\includegraphics[width=1.0\textwidth]{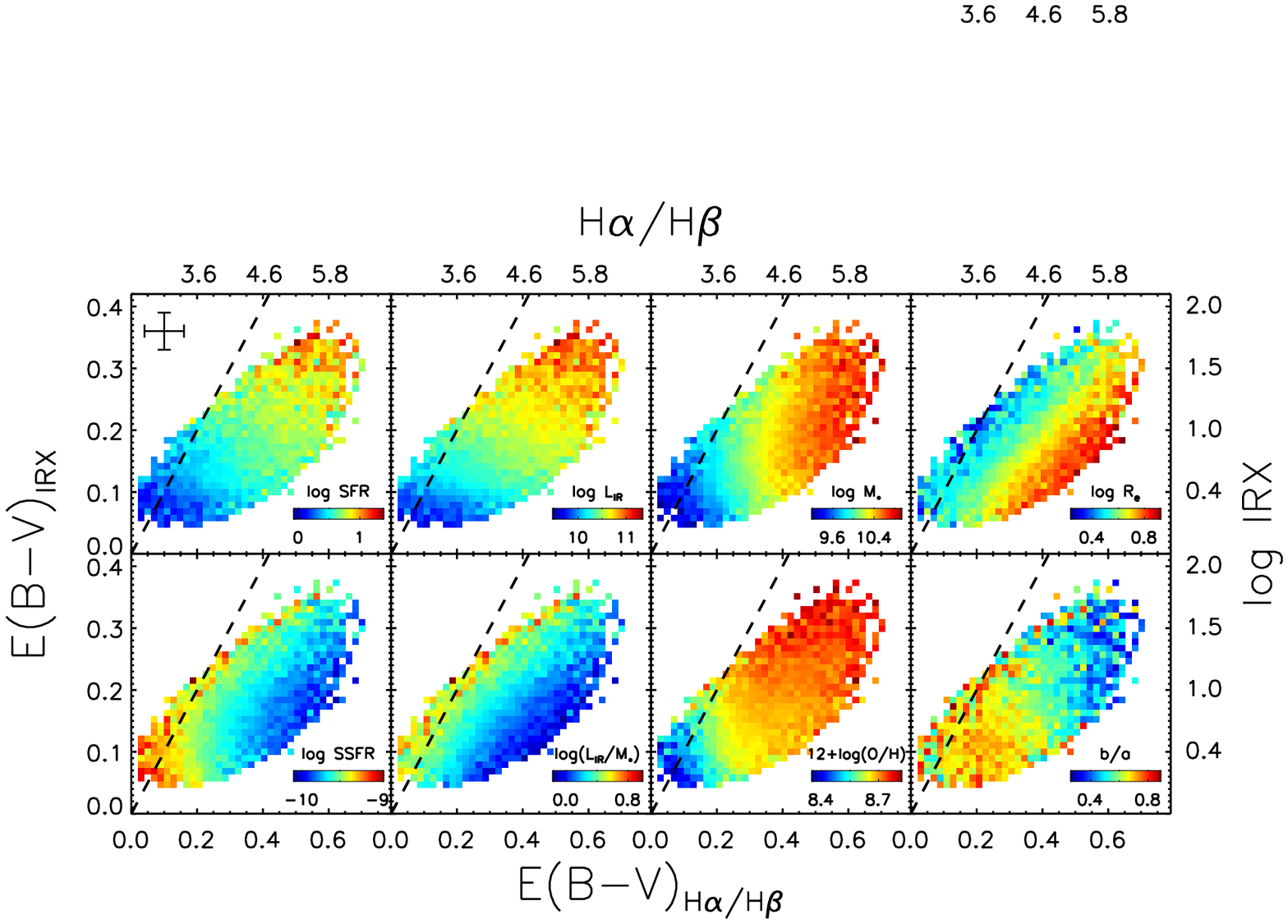}
\caption{Comparison between $\ebvhahb$ and $\ebvirx$ for our sample of 32\,354 local SFGs. The top- and right-axis show the equivalent $\hahb$ and IRX, respectively. We divide each panel into 40$\times$40 subgrids and color code sub-grids using the median of data points in a given parameter. Here we only count the subgrids containing $>$5 SFGs. In these panels from left to right and top to bottom, the parameters used for color coding are  SFR, $L_{\rm IR}$, $M_\ast$, $R_{\rm e}$, SSFR, $L_{\rm IR}/M_\ast$, 12+$\log$(O/H) and $b/a$, respectively.  The dotted lines mark the relation of $\ebvirx=\ebvhahb$ (i.e., $\rebv=1$). The error bars in the top-left panel show the typical observational uncertainties.}
\label{fig:fig1}
\end{figure*}

\section{SAMPLE AND DATA}\label{sec:sec2}

We carry out our investigation of dust attenuation using the sample and data from \citet{Qin2019}, where more details about the sample selection and data extraction can be found. We briefly  recap the most essential aspects here. The sample consists of 32\,354 local SFGs carefully selected from the Sloan Digital Sky Survey Data Release 10 \citep[SDSS DR10,][]{Ahn2014}, the wide surveys by the {\it Galaxy Evolution Explorer} ({\it GALEX}) \citep{Martin2005} and {\it Wide-field Infrared Survey Explorer} ({\it WISE}) All Sky Survey \citep{Wright2010}.   

The cross-matching of SDSS targets with {\it GALEX} and {\it WISE}  photometric catalogs is described in \citet{Salim2016}. The galaxy stellar mass, spectroscopic redshifts and line fluxes of our sample galaxies are taken from the value-added catalog produced by the group from Max Planck Institute of Astrophysics and Johns Hopkins University (MPA/JHU).\footnote{\url{https://www.sdss3.org/dr10/spectro/galaxy_mpajhu.php}} The  sample is restricted to the redshift range from 0.04 to 0.15. We require a fiber coverage fraction ( fiber-to-total stellar mass ratio) $>0.2$, to minimize the potential differences between nuclear and global measurements in metallicity or Balmer decrement \citep{Kewley2005}.\footnote{We verified that a more restrictive cut on fiber coverage fractions does not alter our conclusions.} The emission lines (H$\alpha$, [N\,{\sc ii}], H$\beta$ and [O\,{\sc iii}] with S/N$>$20, 3, 3 and 2, respectively), are used to select SFGs using the `BPT' diagram \citep{Baldwin1981}, and to determine the gas-phase metallicity, as well as to estimate the Balmer decrement. The galaxy  structural parameters, i.e., SDSS $r$-band half-light radius ($R_{\rm e}$) and axial ratio ($b/a$), are taken from the catalog provided by \citet{Simard2011}. We estimate the IR luminosity (8--1000\,$\mu$m) from the {\it WISE} 22\,$\mu$m flux ($>$2\,$\sigma$) using a library of luminosity-dependent IR templates from \citet{Chary2001}. The UV luminosity (1216--3000\,\AA) is calculated by integrating the best-fit galaxy SED template to the observed {\it GALEX} FUV, NUV ($>$3\,$\sigma$) and $u$-band fluxes. The IR-to-UV luminosity ratio is referred to as IRX and SFR is estimated from IR and UV luminosities following \citet{Bell2005}.  

In order to make a quantitative comparison, we convert IRX and $\hahb$ into consistent units of a color excess $\ebvirx$ and $\ebvhahb$, respectively. At first, we derive $A_{\rm FUV}$ from IRX following the relation given by \citet{Buat2005} and then give the color excess with a certain reddening curve $k_{\lambda}$,   
\begin{align}\label{eq:eq1}
& A_{\rm FUV} = -0.0333X^3+0.3522X^2+1.1960X+0.4967,  {\rm and} \nonumber \\
& {\rm E(B-V)}_{\rm IRX}=A_{\rm FUV}/k_{\rm FUV},    
\end{align}
where $X=\log(L_{\rm IR}/L_{\rm FUV})=\log(IRX/1.38)$ and $k_{\rm FUV}=10.22$ for the adopted reddening curve of \citet{Calzetti2000}. 
A factor of 1.38 corrects the FUV luminosity $L_{\rm FUV}$ to the integrated UV luminosity (1216--3000\AA), assuming a SED template of a 100 Myr old stellar population with a constant SFR history. 
For the Balmer decrement (${\rm H\alpha}/{\rm H\beta}$), we estimate $A_{\rm H\alpha}$ and then convert it to a color excess following
\begin{align}\label{eq:eq2}
& A_{\rm H\alpha}=\frac{-2.5k_{\rm H\alpha}}{k_{\rm H\beta}-k_{\rm H\alpha}} \log \left( \frac{2.86}{{\rm H\alpha}/{\rm H\beta}} \right), {\rm and} \nonumber \\ 
& E(B-V)_{\rm H\alpha/H\beta}=A_{\rm H\alpha}/k_{\rm H\alpha}.  
\end{align}
Here 2.86 is the intrinsic $\hahb$ line flux ratio, under the case B  recombination condition with a temperature of $T=10^4\,$K and an electron density of $10^2\,$cm$^{-3}$ \citep{Osterbrock2006}. Adopting the \citet{Calzetti2000} reddening curve, we have $k_{\rm H\beta}=4.60$ and $k_{\rm H\alpha}=3.31$. There are $\sim$60 objects (0.2\% of the sample) with measured $\hahb$$<$2.86, (i.e., $\ebvhahb$$<$0), indicating that the case B condition might not be  an appropriate choice for them. We set $\ebvhahb=0$ for these objects. Then an IRX-to-$\hahb$ color excess ratio of $\rebv\equiv\ebvirx/\ebvhahb$ is introduced to quantify the color excess discrepancy between the two indicators. We note that adopting a Milky Way extinction curve \citep{Cardelli1989} for gas attenuation (i.e., in Equation~\ref{eq:eq2}) as suggested by \citet{Reddy2015} will not alter our conclusions, but bring systematic offset ($\sim$20\%) on $\rebv$.

\section{Results} \label{sec:sec3}
 
We compare $\ebvhahb$ and $\ebvirx$ for 32\,354 local SFGs in Figure~\ref{fig:fig1}. We inspect the number density distribution of galaxies across the diagram that the bulk of galaxies are distributed roughly symmetrically along the locus depicted by the color-coded bins. The SFGs line up along a dust attenuation sequence with more dusty SFGs appearing at higher values of both $\ebvhahb$ and $\ebvirx$. However, the vast majority of our sample galaxies have $\rebv$$<$1. This means that $\hahb$ probes a larger degree of obscuration than IRX.
 
We can see from Figure~\ref{fig:fig1} that SFR, $L_{\rm IR}$, $M_\ast$ and  12+$\log$(O/H) globally increase, and $b/a$ decreases with dust attenuation traced by either  $\ebvhahb$ or $\ebvirx$. These trends are not surprising because more massive SFGs tend to be more dusty, and have higher SFR and metallicity; and more inclined SFGs are more attenuated along the line of sight \citep[see][for a detailed analysis of the relationships between these parameters]{Qin2019}.  


\begin{figure}[htp!]
\centering
\includegraphics[width=1\columnwidth]{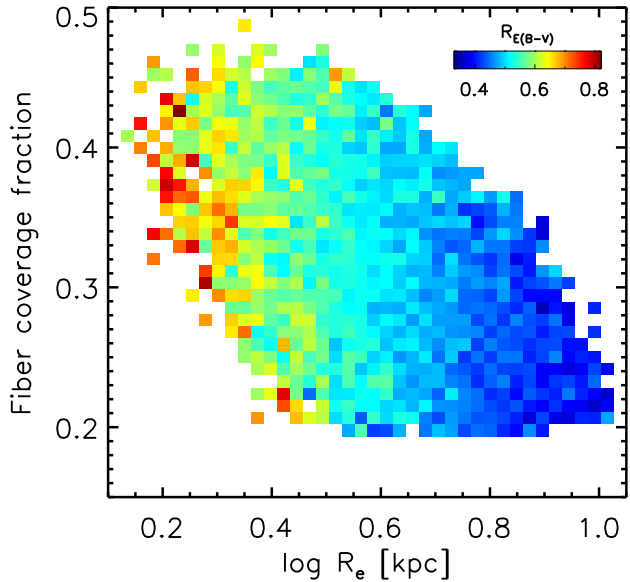}
\caption{$\rebv$ as functions of galaxy half-light radius ($R_{\rm e}$) and fiber coverage fraction. In general galaxies with larger radii have lower fiber coverage fractions.  It is clear that $\rebv$  does not depend on fiber fraction but $R_{\rm e}$.}
\label{fig:fig2}
\end{figure}

 
Figure~\ref{fig:fig1} also shows that $M_\ast$ remains nearly constant with increasing IRX at a fixed $\hahb$ but increases with $\hahb$ at a fixed IRX. This clear dependence of $\ebvhahb$ but not $\ebvirx$ on $M_\ast$ is also reported in \citet{Koyama2019}, using very similar samples.  
 
Strikingly, we find that both SSFR and $R_{\rm e}$ do not change along the dust attenuation sequence defined by $\ebvhahb$ and $\ebvirx$.  Instead, these two parameters exhibit a systematic change roughly perpendicular to the sequence in the sense that data points with lower SSFR and larger $R_{\rm e}$ locate on the right-down side of the sequence. These systematic changes are correlated with $\rebv$. The dependence of $\rebv$ on SSFR is also reported by \citet{Koyama2019}. However,  these authors do not discuss the dependence of $\rebv$ on galaxy size. Figure~\ref{fig:fig2} shows the dependence of $\rebv$ on the physical galaxy size and fiber coverage fraction. It is clear that $\rebv$ decreases with galaxy size at a fixed fiber coverage fraction and there is no clear dependence of $\rebv$ on fiber coverage fraction if galaxy size is fixed. Therefore, we conclude that the correlation between $\rebv$ and galaxy size is not a fiber effect driven by the dust gradient in galaxies \citep[e.g.,][]{Nelson2016}.

We note that  IR luminosity is a good indicator of SFR for normal star-forming and starburst galaxies but no longer a good proxy for the less obscured SFGs (mostly low-mass and low-metallicity). As shown in Figure~\ref{fig:fig1}, $L_{\rm IR}$ (and $L_{\rm IR}/M_\ast$) exhibits a stronger correlation pattern than SFR (and SSFR) in the low attenuation regime. This is understandable because the correlation of IRX is stronger with $L_{\rm IR}$ than with SFR, partially driven by self correlation \citep[see][for discussions]{Qin2019}. For this reason, we make use of SFR instead of $L_{\rm IR}$ in our examination below.

\begin{figure}
\centering
\includegraphics[width=1\columnwidth]{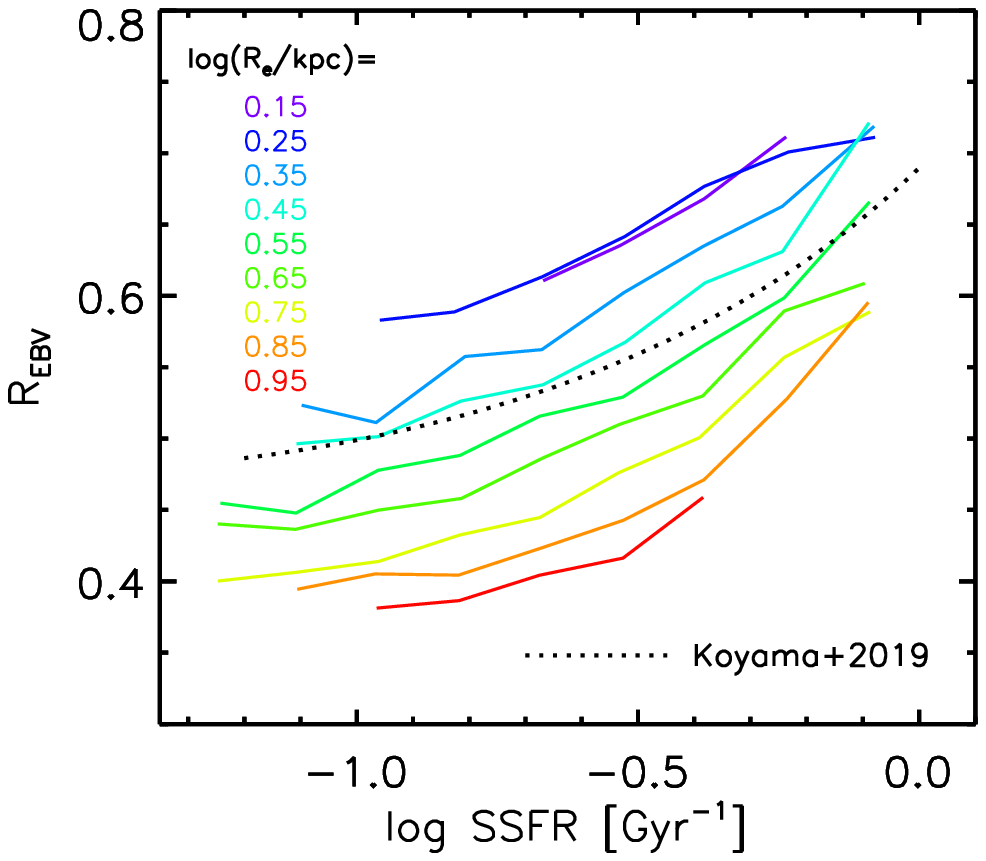}
\includegraphics[width=1\columnwidth]{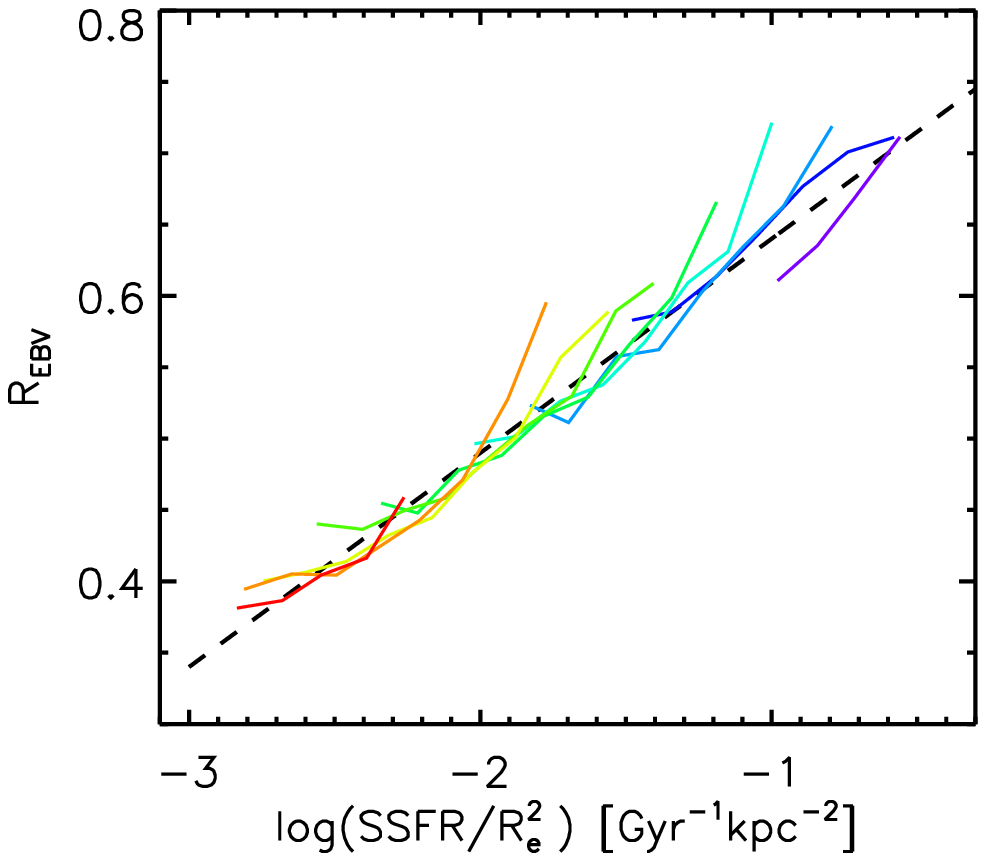}
\caption{Top panel: the SSFR$-$$\rebv$ relation  as a function of $R_{\rm e}$.  Solid lines link the median data points of SSFR$-$$\rebv$ subsamples split by $R_{\rm e}$ (color-coded).  The dotted line is the averaged SSFR$-$$\rebv$ relation given in \citet{Koyama2019}. Bottom panel: the relations between $\rebv$ and $\mussfr$. Color coding is the same as in the top panel. The black dashed line marks the best-fit relation given in Equation~\ref{eq:eq4}. }
\label{fig:fig3}
\end{figure}

\begin{figure*}
\centering
\includegraphics[width=1.0\textwidth]{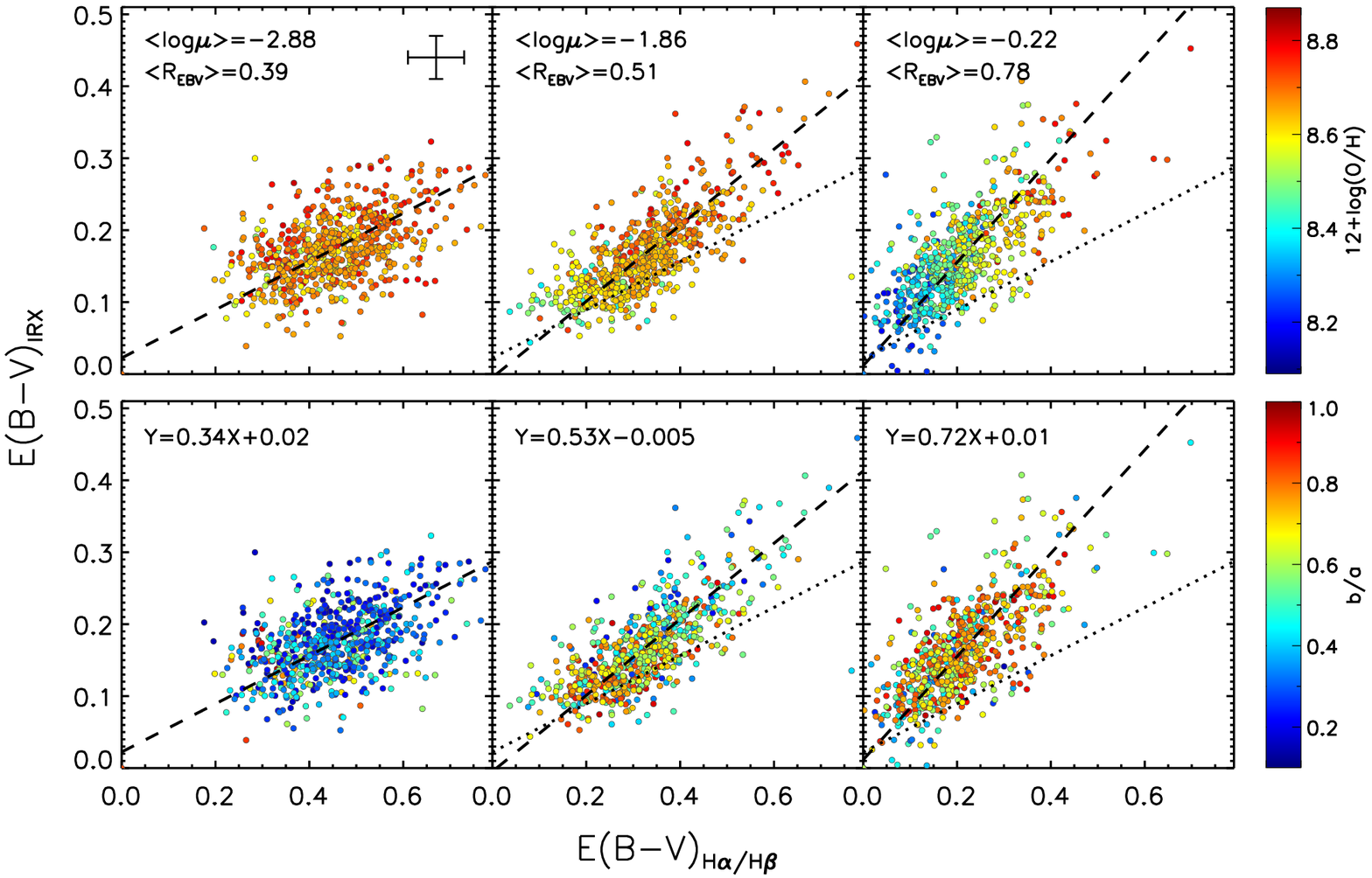}
\caption{ Comparison between  $\ebvhahb$ and $\ebvirx$ for three subsamples of local SFGs with different $\mu$ (=$\mussfr$). The subsamples are selected from 0\%--2\% (left), 49\%--51\% (middle), 98\%--100\% (right) of the cumulative distribution of $\mu$, respectively. The median $\mu$ and $\rebv$ in each bin is labeled. The dashed lines represent the best-fit line $Y=kX+b$ to the data points. The dotted lines in the middle and right panels are the best-fit line in the left panel for comparison. The error bars in the top-left panel show the typical observational uncertainties. Gas-phase metallicity (top panels) and axial ratio (bottom panels) are used to color code data points. At fixed $\log\mu$, the data do not reveal any residual dependence on these parameters.}
\label{fig:fig4}
\end{figure*}

Next we want to quantify the dependence of $\rebv$ on the related galaxy parameters, i.e., SSFR and $R_{\rm e}$ as indicated in Figure~\ref{fig:fig1}. We split the sample into $9\times9$ bins in SSFR and $R_{\rm e}$ with a bin width of 0.15 and 0.1\,dex, respectively. Only those bins having $>60$ galaxies are counted. Here we do not involve the metallicity and axial ratio because they are not correlated with $\rebv$, as shown in Figure~\ref{fig:fig1}. In each (SSFR, $R_{\rm e}$) bin, the median and the dispersion of these parameters as well as $\rebv$ are derived. We show $\rebv$ as a function of SSFR at different $R_{\rm e}$ in the top panel of Figure~\ref{fig:fig3}. It is clear that $\rebv$ increases with SSFR and decreases with $R_{\rm e}$. The dotted line represents the average SSFR-$\rebv$ relation given in \citet{Koyama2019}. However, our finding of the dependence of the SSFR$-$$\rebv$ relation on galaxy size reveals that the connections between IRX and Balmer decrement  can still be mapped more accurately. Considering the power-law relations shown in Figure~\ref{fig:fig3}, we assume that $\rebv$ obeys a formula of 
\begin{equation}\label{eq:eq3}
\rebv=\alpha+\beta\,(\log {SSFR}+\gamma\,\log R_{\rm e}),
\end{equation}   
where $\alpha$, $\beta$ and $\gamma$ are free parameters to be determined via fitting the data points, and the SSFR and $R_{\rm e}$ are given in units of Gyr$^{-1}$ and kpc, respectively. The IDL package {\sc mpfit} \citep{Markwardt2009} is used to fit these binned data points and  $\sigma[\rebv]$ is adopted as the weight for $\rebv$. The $\chi^2$ minimization method is adopted to derive the best-fit parameters, giving $\alpha$=0.79$\pm$0.06, $\beta$=0.15$\pm$0.05 and $\gamma$=$-$1.92$\pm$0.83. Then Equation~\ref{eq:eq3} can be approximately rewritten as 
\begin{equation}\label{eq:eq4}
R_{\rm EBV}=0.79+0.15\times\log\mu, 
\end{equation} 
where $\mu$=$\mussfr$ represents the SSFR surface density in unit of $\rm Gyr^{-1}\,kpc^{-2}$. The bottom panel of Figure~\ref{fig:fig3} shows $\rebv$ as a function of $\mussfr$, largely minimizing the scatter of $\rebv$. \citet{Koyama2019} found that the ``extra attenuation'' (1/$\rebv$) deceases with SSFR. Upon closer inspection, our investigation suggests that the attenuation discrepancy (i.e., $\rebv$) is jointly determined by the SSFR and $R_{\rm e}$ in the form of $\mussfr$. We note that adopting a new form of formula, e.g., second-order polynomial, does not  significantly reduce the scatter. We will discuss the implications of our results in Section~\ref{sec:sec4}. 

We point out that the residuals given in Figure~\ref{fig:fig3} could be further reduced if SFR is replaced with $L_{\rm IR}$ in our analysis. If doing so, i.e., replacing SSFR with $L_{\rm IR}/M_\ast$ in Equation~\ref{eq:eq3}, the best-fit relation would become 
\begin{equation} \label{eq:eq5} 
 R_{\rm EBV}=0.64+0.20\times\log\mu_{\rm IR}, 
\end{equation}
where $\mu_{\rm IR}$=$(L_{\rm IR}/M_\ast)/R_{\rm e}^{1.4}$ in unit of $\rm L_\odot M_\odot^{-1} kpc^{-1.4}$.  Note that the best-fit slope between $\rebv$ and $\log\mu$ increases from 0.15 to 0.20.  Again, the stronger correlation of IRX with $L_{\rm IR}$ than with SFR is partially driven by the self-correlation.  We chose SFR  instead of $L_{\rm IR}$ for further analysis also because SSFR is a direct measure of relative star formation intensity of a galaxy compared to $L_{\rm IR}/M_\ast$. Moreover, the best fit with SSFR yields a correlation of $R_{\rm EBV}$ with SSFR/$R_{\rm e}^2$, which can be physically interpreted as SSFR surface density and thus provides more physical grounds to understand the correlation. 
  
We further examine if the dependence on $\mu$ is sufficient to account for all variations in $\rebv$. Figure~\ref{fig:fig4} shows three subsamples of SFGs with different $\mu$ in the diagram of $\ebvhahb$ versus $\ebvirx$. From the left to the right panel, the median of $\log\mu$ increases from $-$2.88 to $-$0.22 (over a range of $\sim$2.6\,dex), and $\rebv$ increases from 0.39 to 0.78. The middle panel shows a typical value of $\rebv=0.51$ in our sample. The best-fit slopes in different panels approximately represent $\rebv$, as revealed by the fact that these linear fits yield a zero point close to zero. Not only does the slope (or $\rebv$) change, but also the scatter becomes smaller at higher $\mu$. For a given $\mu$ bin, the $\hahb$$-$IRX relation is driven by the overall dust opacity. It is clear from color coding in Figure~\ref{fig:fig4} that the more metal-rich and inclined SFGs generally have higher Balmer decrement and IRX, and show no evidence of bringing scatter on $\rebv$. Moreover, the $\rebv$ residuals from Equation~\ref{eq:eq4} for individual SFGs, defined as $\Delta \rebv=\rebv-R_{\rm EBV,predict}$, do not exhibit any additional correlation with either metallicity or axial ratio, as well as the stellar mass, SFR or galaxy size. This confirms that the input galaxy parameters SSFR and $R_{\rm e}$ fully account for the variation of $\rebv$, without explicit dependence on metallicity and axial ratio. Taken together, we conclude that $\rebv$ is jointly regulated by SSFR and galaxy size, and neither metallicity nor axial ratio significantly influences $\rebv$.

\begin{figure*}
\centering
\includegraphics[width=1\textwidth]{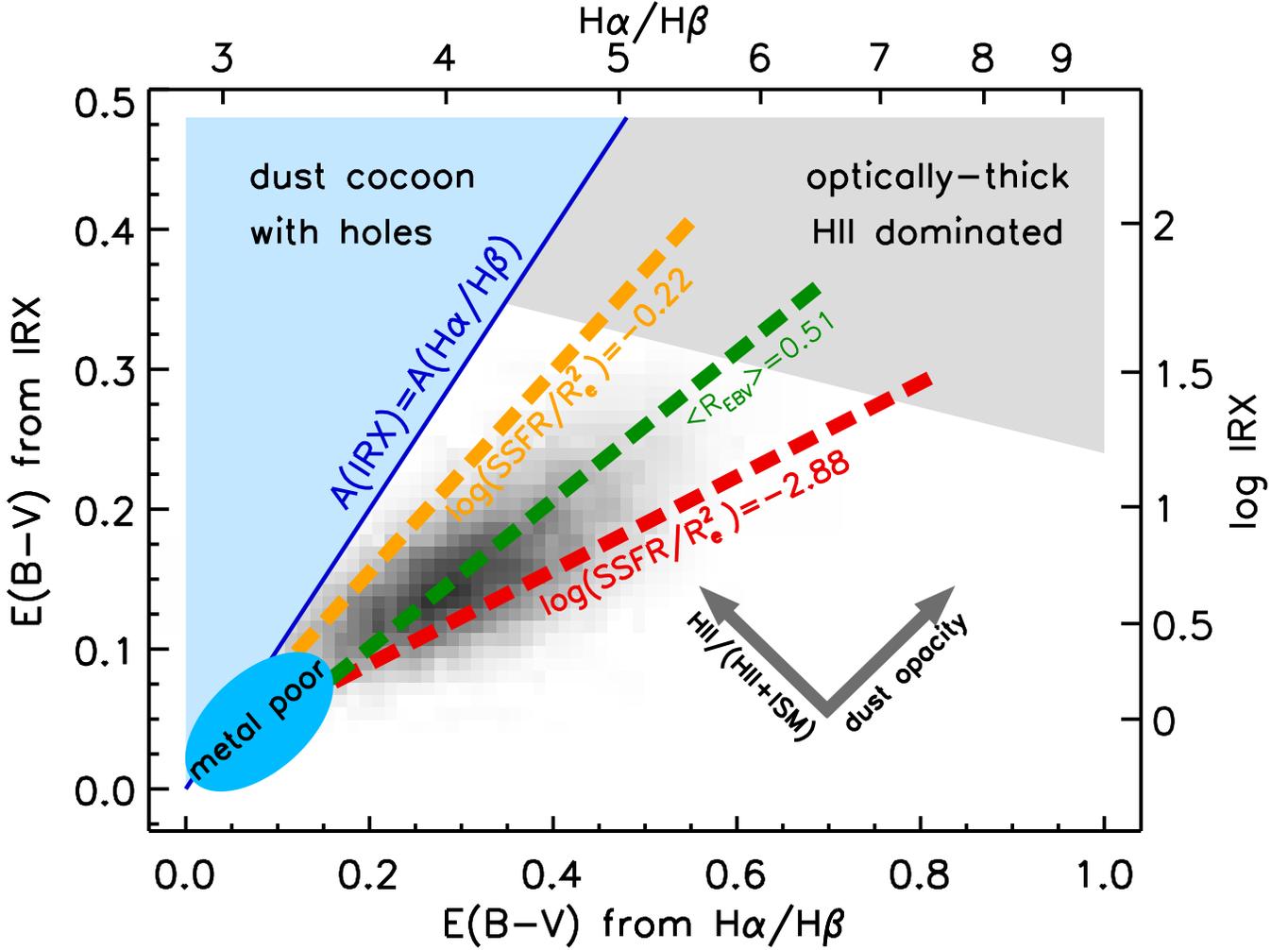}
\caption{Schematic diagram of dust attenuation of SFGs probed by $\hahb$ versus by IRX. The background gray scale represents the distribution of our local 32\,354 SFGs. The yellow, green and red dashed lines represent the best-fit relations to the three subsamples with $\log({\rm SSFR}/R_{\rm e}^2)$ =$-$2.88, $-$1.86 and $-$0.22, respectively, presented in Figure~\ref{fig:fig4}. SFGs with increasing $\mussfr$ form a steeper $\hahb$$-$IRX relation and yield a higher $\rebv$. The increase of $\rebv$ is likely coupled with the increasing fraction of UV emission from \hii\ regions over the total at higher SSFR, as well as the increasing opacity in the diffuse ISM driven by the decreasing galaxy size. When $\rebv$=1, \hii\ regions dominate the parent galaxies and the global dust attenuation is dominated by that in the \hii\ regions. At a fixed $\mussfr$, the $\hahb$$-$IRX relation is driven by the observed dust attenuation that is regulated by SFR compactness, metallicity and inclination.} 
\label{fig:fig5}
\end{figure*}

\section{Discussion} \label{sec:sec4}

We examine the connections between IRX and Balmer decrement in relation to other galaxy parameters, aiming at understanding the discrepancy between the two in probing dust attenuation. Our results reveal that the extra attenuation toward nebular regions, parameterized by the color excess ratio $\rebv$, is jointly regulated by SSFR and galaxy size. SFGs with higher $\mu$=$\mussfr$ form a steeper $\hahb$$-$IRX relation (i.e., higher $\rebv$). At a fixed $\mu$, the position of a galaxy along the $\hahb$$-$IRX relation is driven by the overall dust opacity that is largely affected by metallicity and inclination.  We present a schematic diagram in Figure~\ref{fig:fig5} to demonstrate the key conditions for our results. Our key findings are also expressed in the diagram and we discuss them below. 

On a galaxy scale dust can be seen as homogeneously spreading in the interstellar medium (ISM), which is briefly composed of dense and diffuse components:  short-lived star birth clouds  (or \hii\ regions) and a surrounding diffuse ISM \citep{Charlot2000,Wild2011,Price2014,Reddy2015}. The two-component prescription is often used to explain the attenuation difference between nebulae and stars in galaxies \citep{Charlot2000,Wild2011,Price2014}.  
In principle, Balmer lines come from the \hii\ regions powered by massive O and B stars (lifetime $\sim10^7$\,yr) newly born in dense birth clouds and thus the Balmer decrement probes the attenuation on the dense birth clouds. IRX on the other hand measures the attenuation of the total UV light from young and intermediate-age populations (up to a few $\times10^8$\,yr) processed by dust in \hii\ regions for those stars still embedded in their birth clouds and dust in the diffuse ISM for all stars contributing to the UV emission \citep{Kennicutt2012}. The intermediate-age stellar populations are able to contribute significantly to the total UV radiation \citep{Wuyts2012}. These stars are no longer tightly surrounded by the dense clouds but reside within the diffuse ISM. The UV light from the intermediate-age stars is consequently less attenuated in comparison to that from the \hii\ regions. IRX thus probes the global attenuation averaged over \hii\ regions and the diffuse ISM, reflecting a smaller degree of attenuation than Balmer decrement (i.e., $\rebv<1$).

The difference between nebular and stellar attenuation was suggested to follow E(B-V)$_{\rm star}$/E(B-V)$_{\rm gas}$=0.44 for starburst galaxies \citep{Calzetti2000}. However, our results indicate this ratio to be 0.51  on average. A higher $\rebv$  than 0.44 was also suggested by \citet{Koyama2019}. We point out that the attenuation probed by IRX in this work \citep[and][]{Koyama2019} differs from the stellar continuum attenuation ($A_{\rm star, SED}$) inferred from the rest UV-to-NIR SED modeling in \citet{Calzetti2000}. IRX probes the attenuation mostly on young and intermediate-age stellar populations , while $A_{\rm star, SED}$ measures the attenuation toward stars  of all ages.  In this regard, it is reasonable to find progressively lower attenuation levels from Balmer decrement to IRX to  $A_{\rm star, SED}$ diagnostics.  We thus conclude that the higher values of $\rebv$ reported in this work and by \citet{Koyama2019} compared to the canonical value of 0.44 found by \citet{Calzetti2000} may be expected on physical grounds.

We infer a typical $\rebv = 0.51$ which allows converting $\ebvirx$ to $\ebvhahb$ or vice versa. Importantly, the $\rebv$ increases by a factor of two (0.39--0.78) from low to high SSFR surface density, as shown in Figure~\ref{fig:fig4} (and Figure~\ref{fig:fig5}). The dependence of $\rebv$ on SSFR had been found among SFGs in the local universe \citep{Wild2011,Koyama2019}, as well as at high redshifts \citep{Price2014}. Generally speaking, SSFR is often used as an age indicator for stellar populations in galaxies \citep[e.g.,][]{Brinchmann2004}. With increasing SSFR, the total intrinsic UV emission is increasingly contributed by young massive stars in \hii\ regions, and the dust attenuation discrepancy between IRX and Balmer decrement is then expected to become smaller (see also \citealt{Wild2011}). Moreover, inclusion of galaxy size is able to strengthen this tendency. When the given \hii\ regions and diffuse ISM are distributed over a smaller spatial extent, the dust opacity in the diffuse ISM increases and the attenuation discrepancy between IRX and Balmer decrement becomes smaller. The decrease of galaxy size also leads to a stronger shielding effect such that the UV light from the intermediate-age populations is increasingly attenuated by dust in the \hii\ regions. 
 
On the other hand, the $\hahb$$-$IRX relation is apparently governed by dust opacity once $\rebv$ is fixed.  The change of dust opacity is mainly caused by the variation in the average dust column density along the line of sight. Here we ignore the effects linked to the properties of dust grains (e.g., the size distribution) although these may contribute to the scatter of the relation. To keep the same $\rebv$, the change of dust column density needs to be in lock step in \hii\ regions and the diffuse ISM.  Interestingly, metallicity or inclination (as probed by the projected axial ratio) may play such roles.  It is well known that metallicity essentially controls the dust-to-gas ratio and then the dust opacity \citep{Leroy2011,Remy-Ruyer2014,Qin2019}. The observed dust column density of disk SFGs of a given SSFR surface density can be a function of inclination due to projection \citep{Xiao2012,Qin2019}. Therefore, a group of SFGs with fixed $\mu$ may disperse along the  $\hahb-$IRX relation if their metallicity or axial ratio spreads over a wide range. Indeed,  such patterns related to metallicity and axial ratio are seen in Figure~\ref{fig:fig4}.

We note that our results derived using a sample of mostly normal star-forming galaxies would become invalid in some extreme cases. We present a schematic diagram in Figure~\ref{fig:fig5} to demonstrate the boundary conditions for our results. For metal-poor SFGs,  dust obscuration is low and  insensitive to the star formation compactness \citep{Qin2019}, and the difference between IRX and Balmer decrement becomes negligible. For heavily-obscured galaxies in the gray shaded region, the majority of star formation is totally obscured. The observed Balmer lines come mostly from the less obscured surface regions (i.e., skin effect) and thus the Balmer decrement is no longer a good tracer of dust obscuration of the whole galaxy. Instead, IRX is able to probe the heavily obscured \hii\ regions free from the skin effect. Therefore, the correlation between IRX and Balmer decrement breaks down in the heavily-obscured regime.
On the other hand, in the regime where dust attenuation traced by IRX is higher than that by Balmer decrement ($\rebv>1$), objects tend to contain massive stars (emit UV light) surrounded by the ionized gas (emit Balmer lines) in \hii\ regions. In this case, the optical depth on UV light is expected to be higher than the Balmer lines. There are only few objects in our sample located in this region. 


 
\section{ Summary} \label{sec:sec5}

Using a sample of $\sim$32\,000 local SFGs selected from SDSS, {\it GALEX} and {\it WISE}, we investigate the cause of the different dust attenuations probed by the IRX and $\hahb$ diagnostics. We parameterize their discrepancy by the color excess ratio $\rebv \equiv E(B-V)_{\rm IRX}/E(B-V)_{\hahb}$. Our main results are summarized as follows:  

\begin{enumerate} 
\item The color excess ratio $\rebv$ is jointly controlled by the specific SFR and galaxy size in the form of the SSFR surface density, following $\rebv=0.79+0.15\times\log(\mussfr)$.

\item Neither gas-phase metallicity nor axial ratio is a determining parameter for $\rebv$, although they can affect both  IRX and $\hahb$  in lock step. 

\item We stress that IRX traces the average dust attenuation by the diffuse ISM and partly by \hii\ regions, while all Balmer emission is subject to attenuation from both dust components. We suggest that the SSFR is a measure of the fraction of the intrinsic UV emission from the \hii\ regions to the total (\hii+ISM), while galaxy size influences the dust density (i.e., opacity) in the diffuse ISM.

\end{enumerate}

 We argue that the dust attenuation discrepancy between IRX and Balmer decrement might be universally regulated by the SSFR surface density across cosmic time.  Then a higher $\rebv$ should be expected for the high-$z$ SFGs that are globally higher in SSFR and smaller in size. This is supported by the finding that $\rebv$ increases with  SSFR for high redshift SFGs \citep{Price2014}, and the $\rebv=0.7-0.8$ measured at $z\sim1.6$ \citep{Kashino2013} or $\rebv=0.8$ measured at z$\sim$1 \citep{Broussard2019} are significantly higher than local value. Further efforts are required to examine whether the high-$z$ SFGs follow the empirical $\rebv$ relation observed among local SFGs. 
  
\acknowledgments
We are grateful to the  anonymous referee for her/his useful comments and careful reading of the manuscript. This work is supported by the National Key Research and Development Program of China (2017YFA0402703), NSFC grants (11773076, 11703092), and the Chinese Academy of Sciences (CAS) through a grant to the CAS South America Center for Astronomy (CASSACA) in Santiago, Chile.   ZP acknowledges the support from the Natural Science Foundation of Jiangsu Province (No.BK20161097).

%








\end{document}